\documentstyle[12pt,epsf]{article}
\begin{document}

\title{\Large\bf Underlying Pairing States of High $T_c$ Superconductivity}
\author{{\bf Wei-Min Zhang}\thanks{Talk presented in 1999 Taiwan 
International Conference on Superconductivity (TICS'99) \& The Sixth 
Workshop on Low Temperature Physics (WLTP6) held on August 17-20, 
1999 in Kenting, Taiwan} \\
{\small Department of Physics, National Cheng Kung University, Tainan, 
		Taiwan 701, Republic of China}
\date{}}
\maketitle
\begin{abstract}
{In this talk, I present a microscopic theory I proposed recently
to describe high-$T_c$ superconductivity in cuprates. I show that
coherent pairing states consisting of extended singlet Cooper pairs 
and triplet $\pi$ pairs can manifest both the Mott insulating 
antiferromagnetic order and the $d$-wave superconducting order.
From this configuration of coherent pairing states, I can describe 
both the single electron properties and the low energy collective
excitations in high $T_c$ superconductivity in the same framework. 
The quasiparticle can be derived directly with respect to the coherent 
pairing states. While, the low-lying quantum fluctuations associated 
with spin-wave and charge excitation can be investigated from the 
path integral of the coherent pairing states.}
\end{abstract} 
%
\normalsize

Discovery of high-$T_c$ superconductivity in copper-oxides
reveals very attractive, but also extremely 
complicated, new phenomena in strongly correlated electron 
systems\cite{Dag}. One of the most striking phenomena is 
that cuprates undergo a transition from the Mott insulating 
antiferromagnetic (AF) order to the $d$-wave superconducting 
(dSC) order under dopings. Theorists have tried various 
different mechanisms, such as the theory of resonating valence 
bond (RVB) states of singlet pairs introduced by Anderson a 
decade ago\cite{anderson87} and the SO(5) unified theory of 
AF and dSC order parameters proposed by S. C. Zhang
recently\cite{szhang97}, to explain this metal-insulator
transition. However, a microscopic description of this
AF to dSC transition has not yet been completed.
  
In this talk, I report a microscopic theory I proposed recently
to describe high-$T_c$ superconductivity in cuprates\cite{wzhang99}.
In this theory, I show that
coherent pairing states consisting of extended singlet Cooper pairs 
and triplet $\pi$ pairs can manifest both the Mott 
insulating antiferromagnetic order and the $d$-wave superconducting 
order. At half-filling, this configuration describes an insulating 
AF ordering arisen from the mixing of singlet and triplet pairs. 
Upon doping with holes the $d$-wave pairing gap appears,
and above a certain level cuprates become superconducting. 
In this AF to dSC transition, the nature of no-double-occupied-sites 
and the presence of triplet $\pi$ pairs play an important role.

Based on the square lattice structure of layered copper-oxides,
the cuprate ground state $|\Psi \rangle$ may be 
obtained by projecting out all states of no double occupied 
sites from the following generalized pairing state $|\Phi \rangle$: 
\begin{eqnarray}
	| \Phi \rangle &=& {\prod}'_{\bf k}\exp  \Big\{\eta_1({\bf k})
		c^\dagger_{{\bf k} \uparrow}c^\dagger_{-{\bf k}\downarrow} 
		+ \eta_2({\bf k})c^\dagger_{{\bf k}+{\bf Q} \uparrow}
		c^\dagger_{-{\bf k}+{\bf Q} \downarrow} + \eta_3({\bf k})
		c^\dagger_{{\bf k} \uparrow}c^\dagger_{-{\bf k}+{\bf Q}
		\downarrow}  \nonumber \\
	& & ~~~~~~~~~ + \eta_4({\bf k})
		c^\dagger_{{\bf k} \downarrow}c^\dagger_{-{\bf k}+{\bf Q}
		\uparrow}  + \eta_5({\bf k})c^\dagger_{{\bf k} \uparrow}
		c^\dagger_{-{\bf k}+{\bf Q}\uparrow}+ \eta_6({\bf k})
		c^\dagger_{{\bf k} \downarrow}c^\dagger_{-{\bf k}+{\bf Q}
		\downarrow} - {\rm H.c} \Big\} | 0 \rangle ,	\label{gcs}
\end{eqnarray}
where the production of {\bf k} over the momentum space is restricted 
in the reduced (half) first Brillouin zone, and the complex 
parameters $\eta_i({\bf k})$ are generally {\it link-dependent} 
pairing wave functions. The parity symmetry in the $x-y$ plane 
requires $\eta_i({\bf k}) = \eta_i(-{\bf k})$. In the generalized 
coherent state theory \cite{wzhang90}, $|\Phi \rangle$ is a 
product (over  ${\bf k}$) of the local SO(8)/U(4) coherent 
pairing states \cite{wzhang87}. In fact, Eq.~(\ref{gcs}) is also 
a multi-pair generalization of the standard SU(2) BCS pairing 
state or RVB states (after projection). Explicitly, we may rewrite 
$\eta_1({\bf k})=\eta_s({\bf k})+ \eta_d({\bf k})$, $\eta_2
({\bf k})=\eta_s({\bf k}) - \eta_d({\bf k})$, $\eta_3({\bf k})=
\eta_{\pi_0}({\bf k}) + \eta_\eta({\bf k})$, $\eta_4({\bf k})=
\eta_{\pi_0}({\bf k}) -\eta_{\eta}({\bf k})$, and $\eta_5({\bf k})
=\eta_{\pi_{+}}({\bf k})$, $\eta_6({\bf k})=\eta_{\pi_{-}}({\bf k})$. 
Then, Eq.~(\ref{gcs}) consists of all electron pairs concerned 
in the study of superconductivity. These are the ordinary
$s$-wave Cooper pairs, the extended (i.e. valence bond) singlet 
Cooper pairs [including the extended $s$-wave, $d$-wave and $s+\alpha
d$ pairs], the quasispin $\eta$ pairs, the singlet $p$-wave $\pi$ 
pairs, and finally the triplet $\pi$ pairs. 

However, not all these pairs must be equally important in high 
$T_c$ superconductivity. Indeed, it is no need to include the 
ordinary $s$-wave pairs and the quasispin $\eta$ pairs for the 
AF phase at half-filling.
This is because the ordinary $s$-wave pairs and the quasispin 
$\eta$ pairs describe the conventional $s$-wave SC order parameter 
and the CDW order parameter in the negative-$U$ Hubbard model. 
The ``Shiba" particle-hole transformation on bipartite lattices 
$c^\dagger_{i \uparrow} \rightarrow c^\dagger_{i \uparrow}$, 
$c^\dagger_{i \downarrow} \rightarrow c_{i \downarrow}$ or 
$(-1)^i c_{i \downarrow}$ for $i \in A$ or $B$ sublattice, 
maps these order parameters in the negative-$U$ Hubbard model 
into the staggered AF magnetic order parameters in the 
positive-$U$ Hubbard model at half filling. This leads to 
$\eta_1({\bf k})=-\eta_2({\bf k})$ and $\eta_3({\bf k})=\eta_4({\bf k})$. 
In addition, the cuprate ground states must also be imposed by the 
constraint of no-double-occupied-sites due to the strong repulsive
Coloubm interaction of the on-site electrons. Instead of 
using Gutzwiller projector $P_G$ to remove double occupied 
sites from $|\Phi \rangle$, it is equivalent to require that 
$|\Phi \rangle$ must satisfy the constraint $\langle \Phi |\sum_i (n_{i\uparrow}n_{i\downarrow})| \Phi \rangle = 0$, where summation 
to $i$ is over the lattice sites.  In the mean-field approximation 
of (\ref{gcs}), it is reduced to the global constraint:
\begin{equation}
	{n^2 \over 4N^2} - m^2_s + |\Delta_s|^2
		+ |\Delta_\eta|^2 = 0 ,	\label{gcdo}
\end{equation}
where $n$ is the total electron number,
$m_s$ denotes the 
magnitude of the long-range AF order parameter, and $\Delta_s$ 
and $\Delta_\eta$ are averaged order parameters of the ordinary 
$s$-wave Cooper pairs and the quasispin $\eta$ pairs which {\it 
vanish} ($\Delta_s=\Delta_\eta=0$) under the conditions $\eta_1
({\bf k})=-\eta_2({\bf k})$ and $\eta_3({\bf k})=\eta_4({\bf k})$. 
Therefore, $m^2_s = {n^2 \over 4N^2}$. At half-filling, 
it gives $m_s = 0.5$ which is the same result as in N\'{e}el 
states. The constraint of no-double-occupied-sites also indicates
that the ordinary $s$-wave pairs and the quasispin $\eta$ pairs 
cannot be formed even after hole doping!

One of the importances of taking the cuprate ground states 
as the coherent pairing states (\ref{gcs}) is that it naturally 
introduces a quasiparticle picture with respect to the states,
$c_{\bf k \sigma} | 0 \rangle = 0 \rightarrow \gamma_{\bf k \sigma} 
|\Phi \rangle = 0$, by the generalized Bogoliubov transformuation,
\begin{equation}
	\left[\begin{array}{c}\beta_{\bf k} \\ \beta^\dagger_{\bf -k} 
		\end{array} \right] = \left[\begin{array}{cc} W_{\bf k} 
		& -Z_{\bf k} \\ Z^\dagger_{\bf k} & W^t_{\bf k} 
		\end{array} \right] \left[\begin{array}{c} \alpha_{\bf k} 
		\\ \alpha^\dagger_{\bf -k} \end{array} \right],
\end{equation}
to the Nambu basis: $\alpha_{\bf k} = \{c_{\bf k \uparrow},c_{\bf k 
\downarrow}, c_{\bf k+Q \uparrow},c_{\bf k+Q \downarrow}\}$, and 
$\beta_{\bf k} = \{\gamma_{\bf k \uparrow}, \gamma_{\bf k \downarrow}, 
\gamma_{\bf k+Q \uparrow}, \gamma_{\bf k+Q \downarrow}\}$. Here,
$W^2_{\bf k}+ Z_{\bf k}Z^\dagger_{\bf k} =1$ and $Z_{\bf k}=\eta 
\sin\sqrt{\eta^\dagger \eta}/\sqrt{\eta^\dagger \eta}$ with 
\begin{equation}	\label{etamt}
	\eta ({\bf k}) = {1\over 2} \left[\begin{array}{cccc}0 &
		\eta_1({\bf k}) 
		& \eta_5({\bf k}) & \eta_3({\bf k})\\ -\eta_1({\bf k}) 
	& 0 & \eta_4({\bf k}) & \eta_6({\bf k}) \\ -\eta_5({\bf k}) 
	& -\eta_4({\bf k}) & 0 & \eta_2({\bf k}) \\ -\eta_3({\bf k}) 
	& -\eta_6({\bf k}) & -\eta_2({\bf k}) & 0 \end{array} \right].
\end{equation}
With this quasiparticle picture, one can study various quasiparticle 
properties in high $T_c$ superconductivity, including the single
hole spectrum near the half-filling, the $d$-wave like gap in optimal
dopings and the $d$-wave like spin-gap in underdopings, and also
the evolution of different fermi surfaces recently observed in 
different cuprate componends.

On the other hand, the over-completeness of coherent pairing states 
also span a basic state space for the description of the collective 
excitations observed in cuprate superconductivity, such as the 41 
meV resonance. The partition function can be expressed 
in terms of path integral of the coherent pairing states:
\begin{equation}
	Z(\beta) = \int [d\mu(Z(\tau))] \exp \Big\{ -\int_0^\beta
		d\tau {\cal L} [Z(\tau), \dot{Z}(\tau)] \Big\},
\end{equation}
where
\begin{equation}
	{\cal L} [Z(\tau), \dot{Z}(\tau)] =  \langle 
		\Phi(\tau)| i{d \over d\tau} | \Phi(\tau) \rangle - 
		\langle \Phi(\tau)| H | \Phi(\tau) \rangle 
\end{equation}
is an effective Lagrangian defined on the space of the pairing 
wave-functions, and the $H$ is the Hamiltonian of strong correlated 
electron system, such as the Hubbard model or $t-J$ model Hamiltonian.
This effective Lagrangian contains two terms. The second term is 
just a matrix element of Hamiltonian operator in the coherent pairing 
states. Minimization of this matrix element with respect to the pairing
wave-functions leads to a generalized BCS or RVB theory of the
extended sinplet pairs mixed with triplet pairs \cite{wzhang99}.
The first term is a generalized Berry phase in pairing states, which 
will induce a non-abilian topological gauge field to describe 
quantum fluctuations in collective excitations.
 
Here I only focus on the application of the above general 
theory to the cuprate ground states. Totally spin singlet of 
cuprate ground states requires $\eta_5({\bf k})= - \eta_6({\bf k})$.
Combining with the non-existence of the ordinary $s$-wave pairs
and quasispin $\eta$ pairs by $\eta_1({\bf k})=-\eta_2({\bf k})$, 
$\eta_3({\bf k})=\eta_4({\bf k})$, the generalized Bogoliubov
transformations are restricted by $z_{1\bf k}=-z_{2\bf k}$,
$z_{3\bf k}=z_{4\bf k}$, $z_{5\bf k}=-z_{6\bf k}$, and
$z_{i\bf k}$ is a matrix element of $Z_{\bf k}$ with 
the same form as (\ref{etamt}). These restrictions on the 
pairing wave functions are also necessary for manifestation of 
the $d$-wave pairing symmetry of valence bonds because 
it implies $\eta_i({\bf k+Q}) = - \eta_i({\bf k})$.
Furthermore, because of the spin rotational symmetry, without loss 
of the generality we can define $z_{3\bf k}=z_{d\bf k}, z_{3\bf k}
=z_{\pi\bf k}\cos 2\theta_{\bf k}$ and $z_{5\bf k}=z_{\pi\bf k}
\sin 2\theta_{\bf k}$. 
The AF and dSC gap order parameters in the coherent pairing 
states are given by  
\begin{equation}
	 m_s = {2 \over N} {\sum}'_{\bf k} 
		(z_{d\bf k}z^*_{\pi\bf k}+z^*_{d\bf k}z_{\pi\bf k})~~,~~
	\Delta_d = {1 \over N} {\sum}'_k d({\bf k}) \Big(z^+_{\bf k}
		w^+_{\bf k} +  z^-_{\bf k}w^-_{\bf k} \Big) 
\end{equation}
where $z^\pm_{\bf k}=z_{d\bf k} \pm z_{\pi\bf k}$, and $w^\pm_{\bf k}
=\sqrt{1-|z^\pm_{\bf k}|^2}$. Obviously, without the triplet pairs
(i.e., $z_{\pi\bf k}=0$), the AF order vanishes and the $d$SC order
parameter is reduced to the RVB-type state.  

To be explicit, we may determine the ground state from the $t-J$ 
model. When all electrons in ground states are paired, the 
$t$-term vanishes by the pairing symmetry $z_{1\bf k}
=-z_{2\bf k}$ for any doing.  The next order hoping ($t'$-term) 
has a non-zero expectation value in $|\Phi\rangle$. The ground states 
can be determined by minimizing the $t'-J$ Hamiltonian, 
which leads to the gap equation at zero-temperature,
\begin{equation} \label{gapeq}
	\Delta_{\bf k} = {1 \over N}{\sum}'_{\bf k}V_{\bf kk'}
	{\Delta_{\bf k'} \over 2E_{\bf k'}} ,
\end{equation}
where $\Delta_{\bf k}\equiv \Delta_d d({\bf k}) + \Delta_{es}
\gamma({\bf k})$, 
$V_{\bf kk'}=J[d({\bf k})d({\bf k'})+ \gamma({\bf k})
\gamma({\bf k'})]$ and $E_{\bf k}=\{J^2\Delta^2_{\bf k}+
[\varepsilon({\bf k})-\mu-2(1-\delta)J]^2\}^{1/2}$ with 
$\varepsilon({\bf k})=-4t'\cos k_x\cos k_y$. The fixed 
electron number gives $-{2\over N}{\sum}'_{\bf k}
{\varepsilon({\bf k})-\mu - 2(1-\delta)J \over 
E_{\bf k}}=1-2\delta$, and $\mu$ is the chemical 
potential. The numerical solutions of (\ref{gapeq}) show that 
the $d$-wave gap order parameter appears after dopings 
(but $\delta <0.5$) with maximum peak
$\Delta_d \simeq 0.07 \sim 0.10$ at $\delta \simeq 0.15 
\sim 0.20$ (the typical optimal doping region)
for $t'/J=0.30 \sim 0.20$. This is in good agreement 
with the experimental observations of $d$-wave 
superconducting states in cuprates. It is also 
striking that the extended $s$-wave superconducting
states only emerge in overdoped region of $\delta > 0.5$.
The separation of the $d$-wave states in optimal 
dopings from the extended $s$-wave states in  
overdopings is controlled by the $t'$-term. Here $t'$ must 
be positive. For a negative $t'$, the ordering of $\Delta_d$ 
and $\Delta_{es}$ in terms of $\delta$ will be exchanged, 
which has been excluded by experiments. 
If we let $t'=0$, then $\Delta_d 
= \Delta_{es}$ which have the maximum value 
at doping $\delta=0.5$. This may correspond to the 
symmetry limit of Zhang's SO(5) theory,
although generally the pairing wave functions determined 
here does not form a rigorous global SO(5) group 
structure. The results are plotted in Fig.~1.  
\begin{figure}
  \begin{center}
	\input{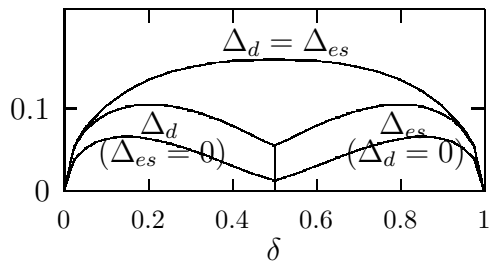}
  \end{center}
  \vskip -0.15in
  \caption{Gap order parameters via dopings at 
zero temperature. The (top) solid line is for 
$t'/J=0$, and the middle and bottom lines 
are for $t'/J=0.2$ and 0.3, respectively.}
\end{figure}

Further applications to the description of dynamical and 
thermal properties of quasiparticles and collective excitations 
measured in cuprate superconductors are in progress.


\end{document}